\documentclass[aps,prb,twocolumn,amsmath,amssymb,superscriptaddress,reprint,longbibliography]{revtex4-2}
\usepackage{amsfonts,amsmath,amssymb,bm}
\usepackage{graphicx,graphics,float}
\usepackage{bm}
\usepackage{amssymb}
\usepackage{colordvi}
\usepackage{graphicx}
\usepackage{color}
\usepackage[colorlinks=true,linkcolor=blue,citecolor=blue,urlcolor=blue]{hyperref}%
\usepackage{hyperref}
\usepackage{comment}
\usepackage{harpoon}
\usepackage{braket}

\begin{document}

\title{Permanent Thermodynamic Mpemba Inversion from Quantum-Jump Feedback}

\author{Jincheng Lu}\email{jinchenglu@usts.edu.cn}   
\affiliation{Key Laboratory of Intelligent Optoelectronic Devices and Chips of Jiangsu Higher Education Institutions, School of Physical Science and Technology, Suzhou University of Science and Technology, Suzhou, 215009, China}
\affiliation{Advanced Technology Research Institute of Taihu Photon Center, School of Physical Science and Technology, Suzhou University of Science and Technology, Suzhou, 215009, China}

\author{Chen Wang}\email{wangchen@zjnu.cn}
\affiliation{Department of Physics, Zhejiang Normal University, Jinhua, Zhejiang 321004, China}

\date{\today}
\begin{abstract}
We show that a common quantum-jump feedback protocol can permanently
reverse the nonequilibrium free-energy ordering of two prescribed
initial states while preserving their Gibbs equilibrium. The
construction exploits collective compensation between monitored
channels: their feedback corrections cancel at equilibrium but remain
active during relaxation. This balance preserves free-energy
monotonicity while allowing the relaxation modes to be reshaped.
We prove that eliminating the complete slow-sector contribution of
the initially higher-free-energy state, while retaining it for the
comparison state, produces distinct asymptotic decay exponents.
Their free-energy ratio then vanishes at long times, guaranteeing
a permanent thermodynamic quantum Mpemba inversion under the same
controlled dynamics. A thermal qutrit realizes the mechanism through
coherence interference and exhibits distinct detection thresholds
for spectral selection and exact slow-sector cancellation under
feedback retuning.
\end{abstract}

\maketitle

\emph{Introduction.---}
Can a common feedback rule reverse the thermal relaxation order of
two fixed quantum states while preserving their Gibbs equilibrium?
The Mpemba effect shows that an initially more distant state can
approach equilibrium faster~\cite{Mpemba1969,Bechhoefer2021,Teza2026}.
In Markovian dynamics, this behavior reflects how initial states
populate the decay modes, with complete suppression of the slowest
contribution producing an exponential separation of relaxation
timescales~\cite{LuRaz2017,Klich2019}.
Controlled observations in colloidal and trapped-ion systems have
established anomalous relaxation as an experimentally accessible
phenomenon~\cite{KumarBechhoefer2020,Kumar2022,Shapira2024,Zhang2025,XuePRL}.
These developments suggest using the Mpemba mechanism to control
relaxation, provided that the initial ordering and the final
equilibrium remain well defined.

Two approaches to such control are initial-state preparation and
Liouvillian engineering.
Unitary preparation can suppress slow-mode amplitudes for a fixed
generator~\cite{Carollo2021,Kochsiek2022,BeatoTeza2026,Caldas26}, whereas
controlled dissipation can reshape the relaxation
pathways~\cite{BaoPRL25,LiuWang2026}.
Dynamics with a prescribed stationary state can also be constructed
through reservoir engineering and measurement-based
feedback~\cite{Guo2025designingopen}.
For thermal relaxation, however, a comparison of trajectories must
account for both energy and entropy.
A quantum state can retain coherence and excess free energy even
when its populations have reached their Gibbs values.
The nonequilibrium free-energy excess therefore provides a
thermodynamic measure that includes both population imbalance and
coherence~\cite{Moroder2024,GooldPRX}.
It remains nonincreasing under Gibbs-preserving dynamical
semigroups~\cite{Lindblad1975,Spohn1978}.
Preserving the Gibbs state during control consequently allows both
trajectories to lose free energy monotonically while their relative
ordering changes.

Quantum-jump feedback provides a concrete way to modify this
relaxation dynamics. Monitoring resolves individual reservoir
events, each of which can trigger a prescribed operation on the
postjump state~\cite{Wiseman1994,PlenioKnight1998,WisemanMilburn2009,
StrasbergPRE}.
The central constraint is to keep these operations dynamically
effective while preserving thermal equilibrium.
For nonzero rank-one jump operators, requiring every feedback
channel to preserve its equilibrium postjump state separately
would leave the unconditional generator unchanged.
Here we exploit collective compensation between channels.
Their feedback corrections cancel at equilibrium, but unequal
jump activities away from equilibrium allow a finite correction
to survive.
The same feedback rule can thus affect different preparations
through their different jump activities without moving the
thermal fixed point.

Within this Gibbs-preserving family, we establish a sufficient
condition for permanent thermodynamic Mpemba inversion.
Feedback removes the complete slow-sector contribution of the
initially higher-free-energy state while retaining a nonzero
contribution for the comparison state.
Their free-energy excesses then decay with different asymptotic
exponents, and their ratio tends to zero.
The reversed ordering therefore holds at every sufficiently late
time under one common controlled dynamics.
A thermal qutrit realizes this selectivity through destructive
interference between initial coherence and coherence generated
from population imbalance.
An exact analysis distinguishes the detection efficiency needed
to make the coherence sector slowest from that needed to cancel
its amplitude, determining when feedback retuning can sustain
the strong effect under imperfect monitoring.

\begin{figure}[htb]
\begin{center}
\centering\includegraphics[width=8.5cm]{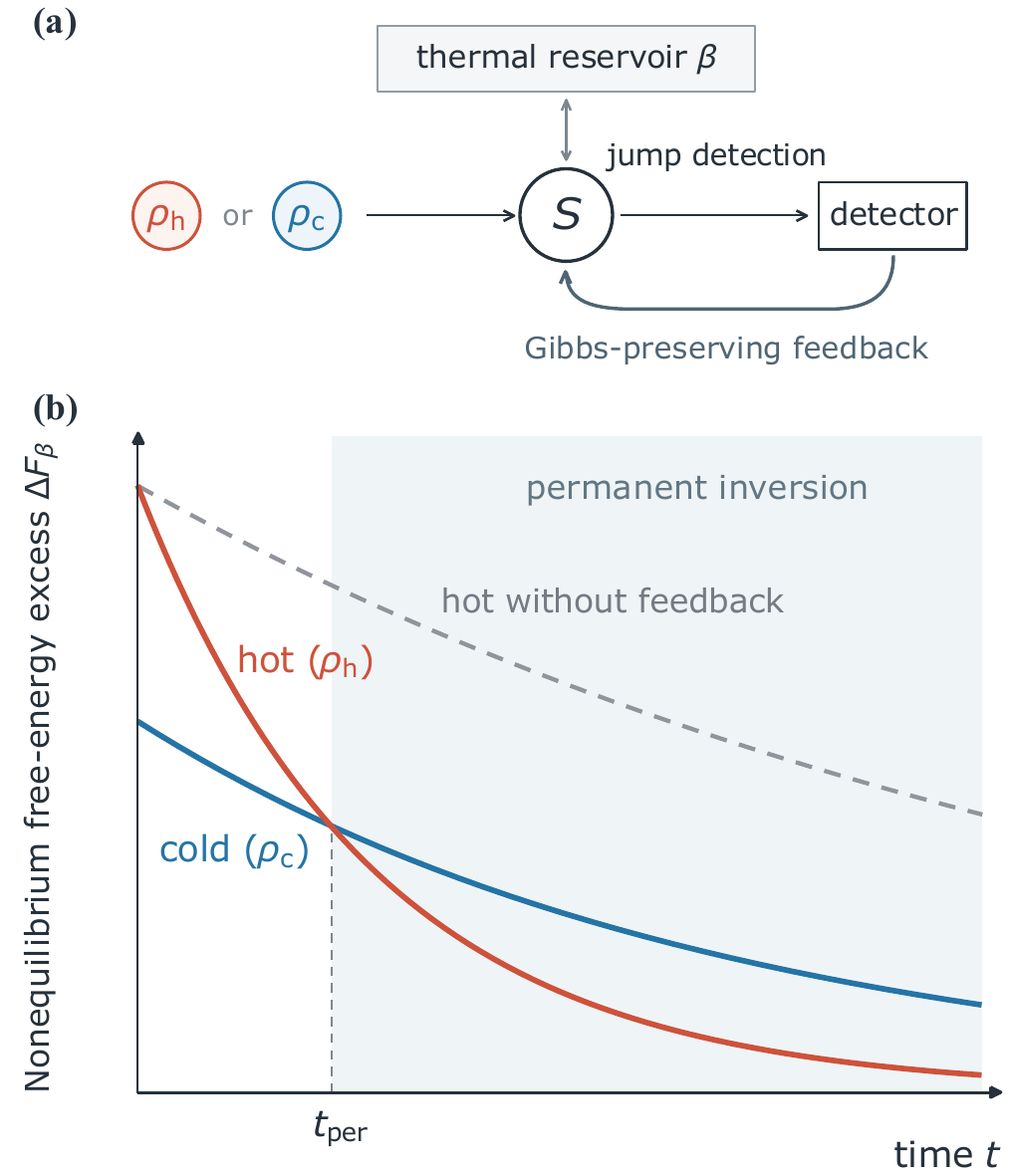}
\caption{
Jump-conditioned feedback and permanent thermodynamic Mpemba inversion.
(a) Either initial state, $\rho_{\rm h}$ or $\rho_{\rm c}$, evolves
under the same feedback protocol. Detected jumps trigger control
pulses that collectively preserve the Gibbs state $\tau_\beta$.
(b) Schematic free-energy relaxation. The initially
higher-free-energy state (red) overtakes the comparison state (blue)
at $t_{\rm per}$, with permanent inversion indicated by shading.
The gray dashed curve shows $\rho_{\rm h}$ without feedback and
never crosses the blue curve. In the illustrated construction,
the comparison trajectory is unchanged by feedback.
}
\label{fig:figure1}
\end{center}
\end{figure}

\emph{Thermodynamic relaxation.---}
We consider a finite-dimensional system with Hamiltonian $H$
weakly coupled to a thermal reservoir at inverse temperature $\beta$.
Its uncontrolled evolution obeys~\cite{Davies1974}
\begin{equation}
 \dot{\rho}(t)
 =
 \mathcal L_0[\rho(t)] ,
 \label{eq_uncontrolled_master}
\end{equation}
where $\mathcal L_0$ is a primitive Davies generator with the unique
attracting Gibbs state $\tau_\beta=e^{-\beta H}/Z$ and
$Z=\operatorname{Tr}(e^{-\beta H})$. Assuming that $\mathcal L_0$
is diagonalizable, relaxation from $\rho_0$ can be expressed as
$\rho(t)-\tau_\beta
=\sum_{j\geq2}a_j(\rho_0)e^{\lambda_jt}r_j$,
where $\lambda_j$ and $r_j$ are the nonstationary eigenvalues and
right eigenoperators, and $a_j(\rho_0)$ are the initial mode
amplitudes. The slowest modes with nonzero amplitudes determine
the long-time approach to equilibrium.
This spectral structure offers two routes to controlling relaxation.
Initial-state preparation changes the mode amplitudes for a fixed
generator~\cite{Carollo2021}, whereas Liouvillian engineering
modifies the generator, allowing the decay rates and mode overlaps
of prescribed initial states to be controlled.

A quantum state can remain out of equilibrium even when its
populations have reached their Gibbs values. If it retains coherence
in the energy basis, it has the equilibrium mean energy but a lower
entropy than $\tau_\beta$. Thermal equilibration therefore involves
both the redistribution of populations and the loss of coherence,
whose energetic and entropic contributions are combined in the
nonequilibrium free energy
$F_\beta(\rho)=\operatorname{Tr}(H\rho)-\beta^{-1}S(\rho)$,
with $S$ the von Neumann entropy.
At fixed $H$ and $\beta$, the Gibbs state minimizes this quantity,
and the excess
$\Delta F_\beta(\rho)
=F_\beta(\rho)-F_\beta(\tau_\beta)
=\beta^{-1}D(\rho\Vert\tau_\beta)$
measures the remaining departure from thermal equilibrium.
Here, $D$ denotes the quantum relative entropy.
For a state with Gibbs populations, this excess comes entirely
from its entropy deficit; more generally, it includes both
population imbalance and coherence.
It thus provides a common thermodynamic measure for comparing
relaxation trajectories~\cite{Moroder2024}.
The excess vanishes only at equilibrium and remains nonincreasing
under a Gibbs-preserving, completely positive and trace-preserving
dynamical semigroup by the data-processing
inequality~\cite{Lindblad1975}.

We consider two prescribed initial states satisfying
$\Delta F_\beta(\rho_{\rm h})
>\Delta F_\beta(\rho_{\rm c})>0$,
where ${\rm h}$ and ${\rm c}$ indicate their initial free-energy
ordering.
A permanent thermodynamic Mpemba inversion occurs when
$\Delta F_\beta(\rho_{\rm h}(t))
<\Delta F_\beta(\rho_{\rm c}(t))$
for every $t>t_{\rm per}$, with finite $t_{\rm per}$.
Our aim is to produce this reversal through a common
jump-conditioned feedback protocol that reshapes the relaxation
modes while keeping both initial preparations and their Gibbs
equilibrium fixed.

\emph{Gibbs-preserving feedback and permanent inversion.---}
Jump-conditioned feedback reshapes relaxation by changing the states
prepared by thermal jumps. Let $L_k$ denote the jump operator for
channel $k$ of the uncontrolled generator $\mathcal L_0$.
Each detected jump is followed immediately by a completely positive
and trace-preserving operation $\mathcal F_k(\boldsymbol\theta)$,
where $\boldsymbol\theta$ collects the feedback parameters.
The same monitoring and feedback rule acts on both initial
preparations, as illustrated in Fig.~\ref{fig:figure1}(a).
Writing $\mathcal J_k[\rho]=L_k\rho L_k^\dagger$, the unconditional
controlled generator is~\cite{StrasbergPRE}
\begin{equation}
 \mathcal L_{\rm fb}(\boldsymbol\theta)
 =
 \mathcal L_0
 +
 \sum_k
 \left[
 \mathcal F_k(\boldsymbol\theta)-\mathcal I
 \right]
 \mathcal J_k ,
 \label{eq_feedback_master}
\end{equation}
where $\mathcal I$ is the identity superoperator.
This generator retains the Lindblad form~\cite{lindblad}.
For unitary feedback,
$\mathcal F_k[X]=U_kXU_k^\dagger$, where $U_k$ is the pulse triggered
by a detected jump in channel $k$, the effective jump operator is
$\widetilde L_k=U_kL_k$.
The identity
$\widetilde L_k^\dagger\widetilde L_k=L_k^\dagger L_k$
ensures that the jump rate
$r_k(\rho)=\operatorname{Tr}(L_k\rho L_k^\dagger)$
is unchanged for a given prejump state, while the postjump state
can be rotated by the feedback.

Preserving the original Gibbs state requires the feedback
corrections to balance at equilibrium. Since
$\mathcal L_0[\tau_\beta]=0$, the necessary and sufficient condition is
\begin{equation}
 \sum_k
 \left[
 \mathcal F_k(\boldsymbol\theta)-\mathcal I
 \right]
 \mathcal J_k[\tau_\beta]
 =
 0 .
 \label{eq_gibbs_preserving_condition}
\end{equation}
We denote the parameters satisfying this condition by
$\mathcal G_\beta$. Gibbs-preserving maps can generate coherence between distinct
energy eigenspaces from energy-diagonal states, a transformation
forbidden by thermal operations~\cite{Faist2015}.  
The balance is collective, allowing individual channels to produce
nonzero corrections that cancel in their sum.
This freedom is essential for the nonzero rank-one jump operators
used below. Their recycling maps have the form
$\mathcal J_k[\rho]=r_k(\rho)\sigma_k$, with a normalized postjump
state $\sigma_k$ independent of $\rho$.
Because $\tau_\beta$ is full rank, $r_k(\tau_\beta)>0$.
Requiring each channel correction to vanish separately at equilibrium
would therefore impose $\mathcal F_k[\sigma_k]=\sigma_k$, making
$[\mathcal F_k-\mathcal I]\mathcal J_k[\rho]=0$ for every state.
Such a requirement would leave the dynamics unchanged,
$\mathcal L_{\rm fb}=\mathcal L_0$.
Collective compensation instead permits feedback to remain active
during relaxation, when the relative jump activities differ from
their equilibrium values.

For $\boldsymbol\theta\in\mathcal G_\beta$, Gibbs invariance and
contractivity of quantum relative entropy ensure that
$\Delta F_\beta$ remains nonincreasing, even when the controlled
generator does not satisfy detailed balance.
We restrict attention to parameters for which
$\mathcal L_{\rm fb}(\boldsymbol\theta)$ is diagonalizable and
$\tau_\beta$ remains its unique attracting stationary state.
Within this family, feedback can change both the decay rates and
the mode amplitudes of a fixed preparation.
In particular, the amplitude
$a_{j,x}^{\boldsymbol\theta}
=\operatorname{Tr}[
(l_j^{\boldsymbol\theta})^\dagger(\rho_x-\tau_\beta)]$
depends on feedback through the left eigenoperator
$l_j^{\boldsymbol\theta}$, with the left and right eigenoperators
chosen biorthonormally.
Thus, changing the feedback can remove a slow contribution without
changing the initial state.

Let $\mathcal P_{\rm s}^{\boldsymbol\theta}$ project onto the
complete slow sector of $\mathcal L_{\rm fb}(\boldsymbol\theta)$,
comprising all nonstationary modes with the largest eigenvalue
real part, including degeneracies and complex-conjugate partners.
A Gibbs-preserving protocol
$\boldsymbol\theta_*\in\mathcal G_\beta$ realizes selective
slow-sector elimination when~\cite{Klich2019,Carollo2021}
\begin{equation}
 \mathcal P_{\rm s}^{\boldsymbol\theta_*}
 [\rho_{\rm h}-\tau_\beta]
 =
 0,
 \qquad
 \mathcal P_{\rm s}^{\boldsymbol\theta_*}
 [\rho_{\rm c}-\tau_\beta]
 \neq
 0 .
 \label{eq_feedback_slow_mode_condition}
\end{equation}
The slow modes remain present in the common generator, but only
the comparison state excites them. The initially higher-free-energy
state consequently approaches equilibrium through faster modes.

This selective cancellation guarantees a permanent thermodynamic
Mpemba inversion. Let $\alpha_{\rm s}<0$ denote the common
eigenvalue real part of the slow sector, and let
$\alpha_{\rm f}<\alpha_{\rm s}$ be the largest real part among
the remaining modes populated by $\rho_{\rm h}$.
The leading state deviations then decay with exponents
$\alpha_{\rm f}$ and $\alpha_{\rm s}$, respectively.
Near the full-rank Gibbs state, relative entropy has a
positive-definite quadratic leading term in the state deviation.
The corresponding free-energy excesses therefore inherit twice
these decay exponents, giving
\begin{equation}
 \frac{
 \Delta F_{\beta,{\rm h}}(t)
 }{
 \Delta F_{\beta,{\rm c}}(t)
 }
 =
 \mathcal O\!\left[
 e^{-2(\alpha_{\rm s}-\alpha_{\rm f})t}
 \right]
 \longrightarrow 0 ,
 \label{eq_permanent_inversion_scaling}
\end{equation}
where
$\Delta F_{\beta,a}(t)
=\Delta F_\beta(\rho_a^{\boldsymbol\theta_*}(t))$
for $a\in\{{\rm h},{\rm c}\}$.
Together with the opposite initial ordering, this limit guarantees
a finite time $t_{\rm per}$ beyond which
$\Delta F_{\beta,{\rm h}}(t)
<\Delta F_{\beta,{\rm c}}(t)$
holds for every subsequent time.
The initially more distant state thus acquires an asymptotic
advantage set by the separation of decay exponents.
Both trajectories continue to approach the same Gibbs state,
while their free-energy ordering remains permanently reversed
[Fig.~\ref{fig:figure1}(b)].

The role of feedback is established by comparing the same
preparations under $\mathcal L_0$.
Define
$\Delta F_{\beta,a}^{(0)}(t)
=\Delta F_\beta(e^{t\mathcal L_0}[\rho_a])$
and let $\alpha_a^{(0)}<0$ be the largest real part among the
uncontrolled modes populated by $\rho_a-\tau_\beta$.
If $\alpha_{\rm h}^{(0)}>\alpha_{\rm c}^{(0)}$, their uncontrolled
free-energy ratio tends to infinity, whereas
Eq.~\eqref{eq_permanent_inversion_scaling} makes the controlled
ratio vanish. Feedback then reverses the asymptotic thermodynamic
ordering.
Its acceleration of $\rho_{\rm h}$ itself follows when
$\alpha_{\rm f}<\alpha_{\rm h}^{(0)}$, since
$\Delta F_{\beta,{\rm h}}(t)/
\Delta F_{\beta,{\rm h}}^{(0)}(t)
=\mathcal O[
e^{-2(\alpha_{\rm h}^{(0)}-\alpha_{\rm f})t}]
\to0$.
The qutrit realization below satisfies both conditions at the
selected operating point and leaves the comparison trajectory
unchanged.

\emph{Finite detection efficiency.---}
Selective slow-sector elimination relies on the feedback applied
after monitored jumps. To determine how much detection is required,
let $\eta\in[0,1]$ denote the probability that a physical jump in
a monitored channel is registered and triggers the prescribed
feedback operation~\cite{CarvalhoPRA,CarvalhoPRA08}. We assume independent
detection losses with the same efficiency for all monitored
channels. A missed jump still changes the system state but receives
no feedback pulse. Averaging over detected and missed events gives
\begin{equation}
 \mathcal L_{\eta,\boldsymbol\theta}
 =
 (1-\eta)\mathcal L_0
 +
 \eta\mathcal L_{\rm fb}(\boldsymbol\theta).
 \label{eq_common_efficiency_generator}
\end{equation}
Thus, $\eta$ continuously connects uncontrolled relaxation
at $\eta=0$ to ideal feedback at $\eta=1$.

Common detection loss preserves the collective balance responsible
for Gibbs invariance: every equilibrium feedback correction is
multiplied by the same factor $\eta$, leaving their cancellation
intact. Consequently, for
$\boldsymbol\theta\in\mathcal G_\beta$, the original Gibbs state
remains stationary and the nonequilibrium free energy remains
nonincreasing at every efficiency. No adjustment of the feedback
parameters is required to preserve this thermal reference.
The dynamical effect of feedback nevertheless weakens as detections
are missed. Although the generator depends linearly on $\eta$,
its eigenvalues and mode overlaps generally do not.

Maintaining the permanent inversion therefore requires controlling
the relaxation modes as the efficiency changes. Feedback angles
that eliminate the slow contribution at ideal detection generally
leave a residual contribution when $\eta<1$. Retuning can compensate
for this change whenever the selective cancellation in
Eq.~\eqref{eq_feedback_slow_mode_condition} remains accessible.
At each efficiency, the same retuned protocol must act on both
initial states. If it restores a strict separation of their
populated decay rates, Eq.~\eqref{eq_permanent_inversion_scaling}
continues to apply: the free-energy ratio tends to zero, and the
inversion remains permanent.

The qutrit realization below resolves two distinct limits on this
control. A spectral-selection threshold $\eta_{\rm sp}$ marks the
onset of access to a coherence-dominated slow sector. A higher
threshold $\eta_c$ marks the onset of exact cancellation of that
sector for $\rho_{\rm h}$. Between these thresholds, feedback can
make coherence the slowest contribution but cannot eliminate its
amplitude from the higher-free-energy state. At and above
$\eta_c$, an appropriately retuned protocol achieves selective
elimination and hence permanent inversion through different
asymptotic decay exponents. The threshold $\eta_c$ therefore
concerns exact strong cancellation; free-energy inversion can
also survive below it when the remaining slow amplitudes favor
$\rho_{\rm h}$. The threshold analysis is presented in Sec.~S5 of the Supplemental
Material~\cite{SM}, where Fig.~S2 illustrates how residual slow-mode
contributions affect the free-energy relaxation.

\begin{figure}[t]
\centering
\includegraphics[width=8.5cm]{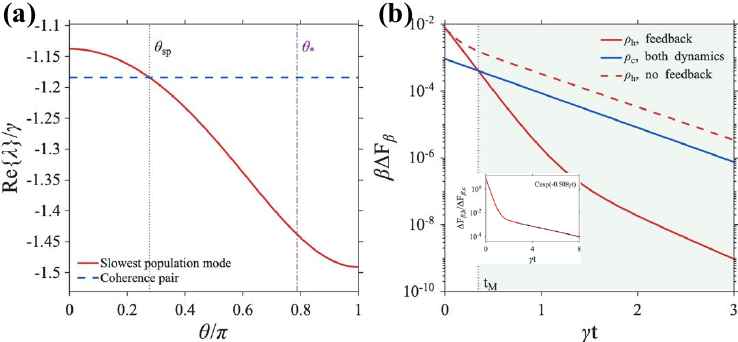}
\caption{Feedback-induced strong thermodynamic Mpemba effect at ideal detection.
(a) Real parts of the leading population eigenvalue and the
coherence pair. The coherence sector becomes slowest beyond
$\theta_{\rm sp}\simeq0.276\pi$; the vertical purple line marks
$\theta_*\simeq0.787\pi$.
(b) Free-energy relaxation at
$(\theta_*,\phi_*)\simeq(0.787\pi,1.904\pi)$.
Selective slow-sector cancellation produces a permanent inversion
after $\gamma t_{\rm M}\simeq0.345$ (shaded region).
The dashed red curve shows uncontrolled relaxation of $\rho_{\rm h}$;
both dynamics give the same $\rho_{\rm c}$ trajectory.
Inset: Free-energy ratio (red) and its independently calculated
asymptotic prediction $C e^{-0.508\gamma t}$ (black dashed).
Parameters are $\beta\gamma=1$, $\omega_1/\gamma=1$,
$\omega_2/\gamma=1.5$, $\Delta p=0.03$,
$\chi_{\rm h}=0.012$, and $\chi_{\rm c}=0.020$.
}
\label{fig:figure2}
\end{figure}

\emph{Qutrit realization.---}
A three-level ladder implements collective Gibbs preservation and
selective slow-sector elimination using two monitored emission
channels. Its Hamiltonian is
$H=\omega_1\lvert1\rangle\langle1\rvert
+(\omega_1+\omega_2)\lvert2\rangle\langle2\rvert$.
The thermal reservoir induces adjacent-level jumps
$L_i^-=\sqrt{\gamma_i^-}\lvert i-1\rangle\langle i\rvert$ and
$L_i^+=\sqrt{\gamma_i^+}\lvert i\rangle\langle i-1\rvert$,
with $\gamma_i^+/\gamma_i^-=e^{-\beta\omega_i}$.
Only the frequency-resolved emission channels are monitored.
Choosing $\gamma_1^-=\gamma$ and
$\gamma_2^-=\gamma e^{\beta\omega_2}$ equalizes their equilibrium
activities, $\gamma_1^-p_1^\beta=\gamma_2^-p_2^\beta\equiv q$,
where $p_i^\beta=\langle i|\tau_\beta|i\rangle$.
Each detected emission triggers the same unitary
$U_{\theta,\phi}$ in the $\lvert0\rangle,\lvert1\rangle$
subspace, with mixing angle $\theta$ and phase $\phi$.
The two equilibrium postjump operators,
$q\lvert0\rangle\langle0\rvert$ and
$q\lvert1\rangle\langle1\rvert$, sum to the identity on this
subspace multiplied by $q$. Their combined contribution is
therefore invariant under the rotation, leaving both feedback
angles available to control relaxation while preserving
$\tau_\beta$. Reservoir engineering provides a route to tailoring
dissipative couplings~\cite{PoyatosPRL}, with quantum
bath engineering demonstrated in superconducting
circuits~\cite{MurchPRL}.

Away from equilibrium, the emission activities generally differ,
and their incomplete cancellation generates
$\lvert0\rangle$--$\lvert1\rangle$ coherence.
We choose $\rho_{\rm h}$ with a population displacement and an
initial coherence, and $\rho_{\rm c}$ with Gibbs populations and
a finite coherence, satisfying
$\Delta F_\beta(\rho_{\rm h})
>\Delta F_\beta(\rho_{\rm c})>0$.
For $\rho_{\rm h}$, feedback controls the amplitude and phase of
the population-induced coherence, allowing its contribution to
the slow mode to cancel that of the initial coherence.
For $\rho_{\rm c}$, the populations remain thermal and generate
no additional coherence, so its trajectory is unchanged.
Once the coherence pair becomes the complete slow sector, this
interference realizes the selective elimination required by
Eq.~\eqref{eq_feedback_slow_mode_condition}.

Figure~\ref{fig:figure2}(a) identifies the spectral regime in which
this cancellation produces a strong Mpemba effect.
Increasing $\theta$ moves the leading population eigenvalue below
the coherence pair at $\theta_{\rm sp}\simeq0.28\pi$.
At the selected point
$(\theta_*,\phi_*)\simeq(0.79\pi,1.90\pi)$, the complete slow
projection of $\rho_{\rm h}$ vanishes, while that of
$\rho_{\rm c}$ remains finite.
Their leading decay exponents are then
$\alpha_{\rm f}\simeq-1.44\gamma$ and
$\alpha_{\rm s}\simeq-1.18\gamma$, respectively.
The free energies cross at $\gamma t_{\rm M}\simeq0.35$
[Fig.~\ref{fig:figure2}(b)].
The inset tests the asymptotic prediction directly:
the propagated free-energy ratio approaches the spectral result
with decay exponent
$2(\alpha_{\rm s}-\alpha_{\rm f})/\gamma\simeq0.51$,
as predicted by Eq.~\eqref{eq_permanent_inversion_scaling}.
Without feedback, $\rho_{\rm h}$ instead retains a population
mode with $\alpha_{\rm h}^{(0)}\simeq-1.14\gamma$, and the
trajectories do not cross over the plotted interval.
Since the entire comparison trajectory is unchanged, these
results demonstrate both permanent inversion and acceleration
of $\rho_{\rm h}$ relative to its own uncontrolled relaxation.

\begin{figure}[t]
\centering
\includegraphics[width=8.5cm]{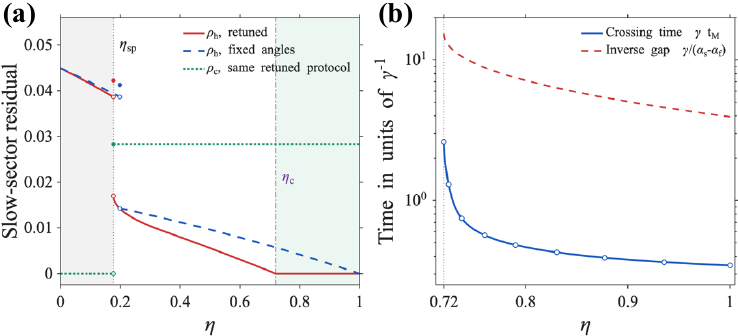}
\caption{
Detection-efficiency dependence of feedback-induced Mpemba relaxation.
(a) Complete slow-sector residual
$\|\mathcal P_{\rm s}^{\eta,\boldsymbol\theta}
[\rho_x-\tau_\beta]\|_{\rm HS}$.
Retuning the feedback (red) restores exact cancellation for
$\rho_{\rm h}$ at $\eta\geq\eta_c\simeq0.72$, while
$\rho_{\rm c}$ retains a finite slow component (green).
The blue dashed curve uses fixed ideal-efficiency angles.
Discontinuities near $\eta\simeq0.2$ reflect changes of the
complete slow sector.
(b) Crossing time $\gamma t_{\rm M}$ (blue) and inverse spectral
gap $\gamma/(\alpha_{\rm s}-\alpha_{\rm f})$ (red dashed)
along the larger-angle exact-cancellation branch.
Increasing efficiency advances the free-energy crossing.
Other parameters are as in Fig.~\ref{fig:figure2}.
}
\label{fig:figure3}
\end{figure}

Figure~\ref{fig:figure3}(a) determines how much detection is
required to retain this mechanism.
Below $\eta_{\rm sp}\simeq0.18$, no feedback angle makes the
coherence pair the complete slow sector.
For $\eta_{\rm sp}<\eta<\eta_c$, this spectral ordering becomes
accessible, but the slow coherence amplitude of $\rho_{\rm h}$
cannot yet be cancelled.
Exact selective elimination first becomes reachable at
$\eta_c\simeq0.72$, allowing approximately $28\%$ missed detections
after retuning.
Keeping the ideal-efficiency angles fixed instead leaves a
nonzero slow residual for $\eta<1$.
The residual discontinuities near $\eta\simeq0.2$ reflect changes
in which modes belong to the complete slow sector.
Thus, spectral selection and amplitude cancellation impose
distinct detection requirements.
The threshold $\eta_c$ concerns exact slow-sector elimination;
permanent inversion can persist below it despite a nonzero slow
residual, as demonstrated in Sec.~S5 and Fig.~S2 of the
Supplemental Material~\cite{SM}.

Figure~\ref{fig:figure3}(b) connects this reachability threshold to
the observable inversion time.
Along the larger-angle exact-cancellation branch,
$\gamma t_{\rm M}$ decreases from approximately $2.6$ at
$\eta_c$ to $0.35$ at ideal detection.
The inverse decay-rate separation decreases alongside it as
feedback opens a larger asymptotic advantage for $\rho_{\rm h}$.
The two timescales remain distinct because the crossing also
depends on the initial mode amplitudes and faster transients.
Detection efficiency therefore controls both access to exact
cancellation and how rapidly its thermodynamic consequence
becomes visible.

Superconducting circuits provide a possible implementation route,
with separate demonstrations of qutrit control, real-time
feedback, and itinerant microwave-photon
detection~\cite{Bianchetti2010,Vijay2012,Lescanne2020}.
The reservoir engineering, emission-detection efficiency, and
feedback latency required to combine these capabilities are
discussed in Sec.~S7 of the Supplemental Material~\cite{SM}.

\emph{Conclusion.---}
We have shown that a common quantum-jump feedback protocol can
enforce a permanent thermodynamic Mpemba inversion, with both
initial preparations and the Gibbs equilibrium held fixed.
Collective compensation between monitored channels preserves
free-energy monotonicity while allowing feedback to reshape the
relaxation modes and their overlaps with each preparation.
Selective cancellation of the complete slow-sector contribution
of the initially higher-free-energy state gives it a faster
asymptotic decay exponent than the comparison state, making the
reversed ordering permanent.
In the qutrit realization, coherence interference accelerates
this state relative to its uncontrolled relaxation while leaving
the comparison trajectory unchanged.
Finite detection efficiency imposes distinct requirements on
spectral selection and exact cancellation, with retuning sustaining
the strong effect above the cancellation threshold.
These results identify collective Gibbs-preserving feedback as a
means of controlling the asymptotic ordering of thermal relaxation
toward a fixed equilibrium.

\emph{Acknowledgments.} We acknowledge support from the National Natural Science Foundation of China (Grant No. 12305050 and No. 12674313), the Natural Science Foundation of Jiangsu Higher Education Institutions of China (Grant No. 23KJB140017), and the Zhejiang Provincial Natural Science Foundation of China under Grant No. LZ25A050001.

\bibliography{RefDQD}

\clearpage

\onecolumngrid  

\tableofcontents

\setcounter{section}{0}
\renewcommand{\thesection}{S\arabic{section}}
\renewcommand{\theequation}{S\arabic{equation}}  
\setcounter{equation}{0}  
\setcounter{figure}{0}
\renewcommand{\thefigure}{S\arabic{figure}}

\section*{Supplemental Material: Permanent Thermodynamic Mpemba Inversion from Quantum-Jump Feedback}

\section{Jump feedback and Gibbs preservation}
\label{sec:S_jump_feedback}

We consider a finite-dimensional system with Hamiltonian $H$ weakly
coupled to a thermal reservoir at inverse temperature $\beta$. Before
feedback is introduced, the system evolves under the Davies generator
\begin{equation}
 \mathcal{L}_{0}[\rho]
 =
 -i[H,\rho]
 +
 \sum_k
 \left(
 L_k\rho L_k^\dagger
 -
 \frac{1}{2}
 \left\{
 L_k^\dagger L_k,\rho
 \right\}
 \right).
 \label{eq:S_uncontrolled_generator}
\end{equation}
We assume that $\mathcal L_0$ is primitive, so that the full-rank Gibbs
state $\tau_\beta=e^{-\beta H}/Z$, with
$Z=\operatorname{Tr}(e^{-\beta H})$, is its unique stationary state
and attracts every physical initial state.

After a detected jump in channel $k$, a completely positive and
trace-preserving map
$\mathcal F_k(\boldsymbol\theta)$ is applied, where
$\boldsymbol\theta$ denotes the feedback parameters. Writing
$\mathcal F_k[X]=\sum_\alpha V_{k\alpha}XV_{k\alpha}^\dagger$ with
$\sum_\alpha V_{k\alpha}^\dagger V_{k\alpha}=\mathbb I_S$, the
controlled jump is represented by the effective operators
$\widetilde L_{k\alpha}=V_{k\alpha}L_k$. Since
$\sum_\alpha\widetilde L_{k\alpha}^\dagger
\widetilde L_{k\alpha}=L_k^\dagger L_k$, feedback changes only the
postjump recycling term. It leaves the no-jump contribution and the
jump probability for a fixed prejump state unchanged.

Defining
$\mathcal J_k[\rho]=L_k\rho L_k^\dagger$, the unconditional feedback
generator is
\begin{equation}
 \mathcal L_{\mathrm{fb}}(\boldsymbol\theta)[\rho]
 =
 \mathcal L_0[\rho]
 +
 \sum_k
 \left[
 \mathcal F_k(\boldsymbol\theta)-\mathcal I
 \right]
 \mathcal J_k[\rho].
 \label{eq:S_feedback_generator}
\end{equation}
The effective jump-operator representation given above shows directly
that this generator has the GKSL form and therefore produces a
completely positive and trace-preserving evolution. Uncontrolled
channels are included by setting $\mathcal F_k=\mathcal I$. For
unitary feedback,
$\mathcal F_k[X]=U_kXU_k^\dagger$, the controlled jump operator is
simply $\widetilde L_k=U_kL_k$. The feedback then rotates the postjump
state without changing the corresponding jump probability.

We now determine when the feedback leaves the original thermal state
unchanged. For each channel, define the unnormalized equilibrium
postjump operator
$X_k^\beta=L_k\tau_\beta L_k^\dagger$. Substituting $\tau_\beta$ into
Eq.~\eqref{eq:S_feedback_generator} gives the Gibbs-preserving
condition
\begin{equation}
 \mathcal L_{\mathrm{fb}}(\boldsymbol\theta)[\tau_\beta]
 =
 \sum_k
 \left[
 \mathcal F_k(\boldsymbol\theta)-\mathcal I
 \right]
 [X_k^\beta]
 =
 0.
 \label{eq:S_gibbs_condition}
\end{equation}
We denote the set of feedback parameters satisfying
Eq.~\eqref{eq:S_gibbs_condition} by
$\mathcal G_\beta$. When local variations inside this set are
considered below, $\boldsymbol\xi$ denotes a set of independent
coordinates on $\mathcal G_\beta$.

Equation~\eqref{eq:S_gibbs_condition} is a collective condition.
Individual feedback maps need not preserve their corresponding
equilibrium postjump operators. A modification produced in one
channel may instead be compensated by the feedback corrections from
the other channels. This collective cancellation provides more
control freedom than imposing
$\mathcal F_k[X_k^\beta]=X_k^\beta$ separately for every channel.

Gibbs preservation guarantees that $\tau_\beta$ remains stationary,
but does not by itself guarantee that it is the unique attracting
state. We therefore restrict the following analysis to parameters in
$\mathcal G_\beta$ for which
$\mathcal L_{\mathrm{fb}}(\boldsymbol\theta)$ has $\tau_\beta$ as its
unique globally attracting stationary state. Within this admissible
family, feedback can reshape the decay rates and eigenoperators
without changing the common thermal endpoint.

\emph{Generality of the collective compensation mechanism}

The construction in the main text is illustrated with a thermal
qutrit, but the mechanism itself does not rely on the three-level
structure. What it requires is a set of monitored jump channels whose
feedback corrections cancel collectively at the Gibbs state, while
remaining active away from it. We make this condition precise and
discuss the minimal resources it demands.

Recall the Gibbs-preserving condition of Eq.~\eqref{eq:S_gibbs_condition},
for a single monitored channel, this reduces to
$\mathcal F_1[X_1^\beta]=X_1^\beta$ with
$X_1^\beta=L_1\tau_\beta L_1^\dagger$. Since
$\mathcal F_1$ is CPTP and $X_1^\beta$ is positive, this condition
forces $\mathcal F_1$ to leave the equilibrium postjump state
invariant. For a rank-one channel, the argument in Sec.~\ref{sec:S_jump_feedback}
then implies $[\mathcal F_1-\mathcal I]\mathcal J_1[\rho]=0$ for every
state $\rho$, so the feedback becomes dynamically inert. A single
monitored rank-one channel therefore cannot support nontrivial
Gibbs-preserving control.

Two channels already suffice. Suppose the equilibrium postjump
operators satisfy
\begin{equation}
 \mathcal F_1[X_1^\beta]-X_1^\beta
 =
 -\left(
 \mathcal F_2[X_2^\beta]-X_2^\beta
 \right)
 \neq 0.
 \label{eq:S_generality_two_channel}
\end{equation}
Then Eq.~(S3) holds, but
neither channel is individually Gibbs preserving. Away from
equilibrium, the two channel activities
$\nu_k(\rho)=\operatorname{Tr}\mathcal J_k[\rho]$ generally differ,
so the cancellation is incomplete and a finite feedback correction
survives. This is precisely the mechanism exploited by the qutrit.
The equal-activity condition
$\gamma_1^-p_1^\beta=\gamma_2^-p_2^\beta$ used there is a convenient
way to realize Eq.~\eqref{eq:S_generality_two_channel}, but it is
not the only one. Any pair of spectrally resolved channels whose
equilibrium postjump operators are related by a feedback-dependent
rotation can satisfy the same collective balance.

The mechanism therefore applies whenever the following ingredients
are available: (i) at least two monitored channels with distinct
postjump states, (ii) a feedback operation that acts differently on
these postjump states, and (iii) a parameter regime in which the
collective correction vanishes at $\tau_\beta$ but not away from it.
Condition (iii) is the nontrivial one, because it requires the
equilibrium activities of the channels to be balanced. In the
qutrit this balance is arranged by choosing
$\gamma_2^-=\gamma_1^-e^{\beta\omega_2}$; in a general system it
would need to be verified for the given rates and temperature.

We emphasize the difference from two related control strategies.
Initial-state rotations~\cite{Carollo2021,Kochsiek2022} suppress
selected Liouvillian modes by rotating the prepared state before
relaxation begins. The rotation is static and must be optimized
separately for each initial state. Dissipative-pathway
engineering~\cite{LiuWang2026} modifies the jump operators themselves,
which changes the thermal fixed point unless additional constraints
are imposed. The present mechanism instead acts dynamically through
the measurement record, uses the same feedback rule for both initial
states, and preserves the original Gibbs state by collective
cancellation. These three features together distinguish it from
existing mode-control schemes.

Finally, we note the limitations. If the monitored channels have
unequal detection efficiencies, the collective balance condition
acquires efficiency weights ,
and the feedback parameters must be retuned accordingly. If the
channel activities cannot be balanced at equilibrium, the Gibbs
state drifts under feedback and the free-energy contraction
argument no longer applies. Whether a given platform admits a
balanced pair of monitored channels is therefore a model-dependent
question, but it is a question of parameter choice rather than of
principle.

\section{Free-energy contraction under Gibbs-preserving feedback}
\label{sec:S_free_energy_contraction}

For fixed $H$ and $\beta$, the nonequilibrium free energy is defined as
$F_\beta(\rho)=\operatorname{Tr}(H\rho)-\beta^{-1}S(\rho)$, where
$S(\rho)=-\operatorname{Tr}(\rho\ln\rho)$. Its equilibrium value is
$F_{\mathrm{eq}}=-\beta^{-1}\ln Z$, with
$Z=\operatorname{Tr}(e^{-\beta H})$. Since
$\ln\tau_\beta=-\beta H-\ln Z$, the quantum relative entropy with
respect to the Gibbs state satisfies
$D(\rho\Vert\tau_\beta)
=\operatorname{Tr}(\rho\ln\rho)
+\beta\operatorname{Tr}(H\rho)+\ln Z
=\beta[F_\beta(\rho)-F_{\mathrm{eq}}]$.
The free-energy excess can therefore be written as
\begin{equation}
 \Delta F_\beta(\rho)
 \equiv
 F_\beta(\rho)-F_{\mathrm{eq}}
 =
 \beta^{-1}
 D(\rho\Vert\tau_\beta),
 \label{eq:S_free_energy_excess}
\end{equation}
where
$D(\rho\Vert\sigma)
=\operatorname{Tr}[\rho(\ln\rho-\ln\sigma)]$
is the quantum relative entropy. Because the finite-temperature Gibbs
state is full rank, its support contains that of every system state, so
$D(\rho\Vert\tau_\beta)$ is finite. Klein's inequality further gives
$D(\rho\Vert\tau_\beta)\geq0$, with equality if and only if
$\rho=\tau_\beta$. Hence $\Delta F_\beta$ is a nonnegative measure of
the thermodynamic distance from the prescribed Gibbs state.

For a fixed feedback parameter
$\boldsymbol\theta\in\mathcal G_\beta$, let
$\Lambda_t^{\boldsymbol\theta}
=\exp[t\mathcal L_{\mathrm{fb}}(\boldsymbol\theta)]$
denote the feedback-controlled dynamical map. Since
$\mathcal L_{\mathrm{fb}}$ has the GKSL form,
$\Lambda_t^{\boldsymbol\theta}$ is completely positive and trace
preserving for every $t\geq0$. Its time independence also implies the
semigroup property
$\Lambda_{t+s}^{\boldsymbol\theta}
=\Lambda_s^{\boldsymbol\theta}\Lambda_t^{\boldsymbol\theta}$.
Consequently, if
$\rho(t)=\Lambda_t^{\boldsymbol\theta}[\rho(0)]$, then
$\rho(t+s)=\Lambda_s^{\boldsymbol\theta}[\rho(t)]$.

The defining condition
$\boldsymbol\theta\in\mathcal G_\beta$ gives
$\mathcal L_{\mathrm{fb}}(\boldsymbol\theta)[\tau_\beta]=0$.
Expanding the exponential map shows that
$\Lambda_s^{\boldsymbol\theta}[\tau_\beta]
=\tau_\beta$, since every term containing a positive power of
$\mathcal L_{\mathrm{fb}}$ vanishes when acting on $\tau_\beta$.
Thus, both $\rho(t)$ and the reference state $\tau_\beta$ may be
propagated through the same CPTP map
$\Lambda_s^{\boldsymbol\theta}$. Applying the data-processing
inequality,
$D(\Phi[\rho]\Vert\Phi[\sigma])
\leq D(\rho\Vert\sigma)$ for every CPTP map $\Phi$, yields
\begin{equation}
 \begin{aligned}
 \Delta F_\beta(\rho(t+s))
 &=
 \beta^{-1}
 D\!\left(
 \rho(t+s)\Vert\tau_\beta
 \right)
 \\
 &=
 \beta^{-1}
 D\!\left(
 \Lambda_s^{\boldsymbol\theta}[\rho(t)]
 \,\middle\Vert\,
 \Lambda_s^{\boldsymbol\theta}[\tau_\beta]
 \right)
 \\
 &\leq
 \beta^{-1}
 D\!\left(
 \rho(t)\Vert\tau_\beta
 \right)
 =
 \Delta F_\beta(\rho(t)),
 \qquad s\geq0.
 \end{aligned}
 \label{eq:S_free_energy_contraction}
\end{equation}
The first equality uses
$\Delta F_\beta=\beta^{-1}D(\,\cdot\,\Vert\tau_\beta)$, the second uses
the semigroup property together with Gibbs preservation, and the
inequality is the contraction of quantum relative entropy under CPTP
maps. Equation~\eqref{eq:S_free_energy_contraction} therefore shows
that Gibbs preservation is sufficient for the nonequilibrium free
energy defined by the original $H$ and $\beta$ to remain
nonincreasing under feedback. Detailed balance of the controlled
generator is not required.

Gibbs preservation alone does not imply strict contraction. The
data-processing inequality may be saturated for particular states or
over finite time intervals, and the controlled generator may in
principle possess additional stationary states or nondecaying
subspaces. We therefore restrict the following analysis to feedback
parameters for which $\tau_\beta$ remains the unique attracting
stationary state. Under this assumption,
$\rho(t)\rightarrow\tau_\beta$ and hence
$\Delta F_\beta(\rho(t))\rightarrow0$ for every physical initial
state, although the decay need not be strictly monotonic at every
instant.

The monotonic contraction of each individual trajectory does not
exclude a Mpemba inversion because it places no restriction on the
relative decay rates of two different initial states. Two states may
both approach $\tau_\beta$ monotonically while their free-energy
ordering changes. In particular, feedback can eliminate the complete
slow contribution of the initially higher-free-energy state while
leaving a nonzero slow component in the comparison state. The former
then decays asymptotically faster, allowing their free energies to
cross and remain inverted at sufficiently long times.

The contraction established here concerns the reduced state of the
controlled system. The direct energy exchanged with the feedback
operations is treated separately below.

\section{Spectral condition for permanent free-energy inversion}
\label{sec:S_spectral_inversion}

Fix an admissible feedback parameter
$\boldsymbol\theta_*\in\mathcal G_\beta$ and consider the corresponding
generator
$\mathcal L_{\mathrm{fb}}
=\mathcal L_{\mathrm{fb}}(\boldsymbol\theta_*)$.
We assume that it is diagonalizable and that $\tau_\beta$ is its
unique attracting stationary state. Its right and left eigenoperators
are defined by
\begin{equation}
 \mathcal L_{\mathrm{fb}}[r_j]
 =
 \lambda_jr_j,
 \qquad
 \mathcal L_{\mathrm{fb}}^\dagger[l_j]
 =
 \lambda_j^*l_j,
 \qquad
 \operatorname{Tr}(l_j^\dagger r_m)
 =
 \delta_{jm}.
 \label{eq:S_biorthogonal_eigenoperators}
\end{equation}
The stationary mode can be chosen as
$\lambda_1=0$, $r_1=\tau_\beta$, and $l_1=\mathbb I_S$.
All remaining eigenvalues satisfy
$\operatorname{Re}\lambda_j<0$.

For an initial state $\rho_x$, where
$x\in\{{\rm h},{\rm c}\}$, define
$\delta\rho_x=\rho_x-\tau_\beta$ and the mode amplitude
$a_{j,x}=\operatorname{Tr}(l_j^\dagger\delta\rho_x)$. Its deviation
from equilibrium evolves as
\begin{equation}
 \delta\rho_x(t)
 =
 \rho_x(t)-\tau_\beta
 =
 \sum_{j\geq2}
 a_{j,x}e^{\lambda_jt}r_j.
 \label{eq:S_spectral_expansion}
\end{equation}
The real part of $\lambda_j$ determines the decay rate, while its
imaginary part produces an oscillation.

Let
$\alpha_{\rm s}=\max_{j\geq2}\operatorname{Re}\lambda_j<0$
denote the slowest nonstationary spectral real part and define the
complete slow index set by
$\mathcal S=\{j\geq2:
\operatorname{Re}\lambda_j=\alpha_{\rm s}\}$.
The corresponding spectral projector is
\begin{equation}
 \mathcal P_{\rm s}[X]
 =
 \sum_{j\in\mathcal S}
 r_j\operatorname{Tr}(l_j^\dagger X).
 \label{eq:S_slow_projector}
\end{equation}
The set $\mathcal S$ includes every mode with the same slow decay
rate. In particular, both members of a complex-conjugate pair must be
included. The projector $\mathcal P_{\rm s}$ is an algebraic spectral
projector and need not represent a physical quantum channel.

We consider two prescribed initial states satisfying
$\Delta F_\beta(\rho_{\rm h})
>\Delta F_\beta(\rho_{\rm c})>0$.
A feedback-induced strong Mpemba point obeys
\begin{equation}
 \mathcal P_{\rm s}
 [\rho_{\rm h}-\tau_\beta]
 =
 0,
 \qquad
 \mathcal P_{\rm s}
 [\rho_{\rm c}-\tau_\beta]
 \neq
 0.
 \label{eq:S_strong_mpemba_condition}
\end{equation}
The first condition removes the complete slow sector of the initially
higher-free-energy state, whereas the second ensures that the
comparison state retains a finite slow contribution. The independent
real components of these projections will be denoted by
$\boldsymbol g_x\in\mathbb R^{d_{\rm s}}$ in the controllability
analysis below.

We now connect the spectral condition to the thermodynamic free
energy. Since $\tau_\beta$ is full rank, the relative entropy admits a
quadratic expansion for every sufficiently small Hermitian traceless
deviation $\delta\rho$,
\begin{equation}
 \Delta F_\beta(\tau_\beta+\delta\rho)
 =
 \frac{1}{2\beta}
 \operatorname{Tr}
 \left[
 \delta\rho\,
 \mathcal T_{\tau_\beta}[\delta\rho]
 \right]
 +
 \mathcal O(\|\delta\rho\|^3),
 \label{eq:S_free_energy_local_expansion}
\end{equation}
where $\mathcal T_{\tau_\beta}$ is the Fr\'echet derivative of the
matrix logarithm at $\tau_\beta$, defined by
$\mathcal T_{\tau_\beta}[X]
=\left.d\ln(\tau_\beta+\epsilon X)/d\epsilon\right|_{\epsilon=0}$.
In the eigenbasis of $\tau_\beta$, the coefficient multiplying
$|X_{mn}|^2$ in the associated quadratic form is
$(\ln p_m^\beta-\ln p_n^\beta)/(p_m^\beta-p_n^\beta)$, with the
continuous value $1/p_n^\beta$ when the two probabilities coincide.
Every coefficient is positive. The quadratic form is therefore
positive definite and equivalent to the Hilbert--Schmidt norm in
finite dimension.

Let $\alpha_{\rm f}<\alpha_{\rm s}$ be the largest real part among the
remaining modes with nonzero amplitude in $\rho_{\rm h}$ after the
slow sector has been eliminated. Equation~\eqref{eq:S_strong_mpemba_condition}
then gives
\begin{subequations}
\label{eq:S_state_asymptotics}
\begin{align}
 \delta\rho_{\rm h}(t)
 &=
 \mathcal O
 \left(
 e^{\alpha_{\rm f}t}
 \right),
 \label{eq:S_higher_state_asymptotics}
 \\
 \delta\rho_{\rm c}(t)
 &=
 e^{\alpha_{\rm s}t}X_{\rm s}(t)
 +
 o
 \left(
 e^{\alpha_{\rm s}t}
 \right).
 \label{eq:S_comparison_state_asymptotics}
\end{align}
\end{subequations}
Here, $X_{\rm s}(t)$ contains the bounded oscillations within the slow
sector. Because the slow sector is finite dimensional and
diagonalizable, a nonzero initial slow projection implies the
existence of constants $c_-,c_+>0$ such that
$c_-\leq\|X_{\rm s}(t)\|_{\rm HS}\leq c_+$.
Complex slow modes therefore cannot remove the asymptotic slow weight
through destructive interference at isolated late times.

Substitution of Eq.~\eqref{eq:S_state_asymptotics} into the quadratic
free-energy expansion gives
\begin{subequations}
\label{eq:S_free_energy_asymptotics}
\begin{align}
 \Delta F_\beta(\rho_{\rm h}(t))
 &=
 \mathcal O
 \left(
 e^{2\alpha_{\rm f}t}
 \right),
 \label{eq:S_higher_free_energy_asymptotics}
 \\
 \Delta F_\beta(\rho_{\rm c}(t))
 &=
 \Theta
 \left(
 e^{2\alpha_{\rm s}t}
 \right).
 \label{eq:S_comparison_free_energy_asymptotics}
\end{align}
\end{subequations}
The symbol $\Theta$ means that the second quantity is bounded above
and below by positive multiples of
$e^{2\alpha_{\rm s}t}$ at sufficiently long times. The cubic
remainder in Eq.~\eqref{eq:S_free_energy_local_expansion} decays faster
than the quadratic contribution and does not affect these exponents.

Since $\alpha_{\rm f}<\alpha_{\rm s}<0$, the ratio of the two
free-energy excesses satisfies
\begin{equation}
 \frac{
 \Delta F_\beta(\rho_{\rm h}(t))
 }{
 \Delta F_\beta(\rho_{\rm c}(t))
 }
 =
 \mathcal O
 \left[
 e^{-2(\alpha_{\rm s}-\alpha_{\rm f})t}
 \right]
 \longrightarrow
 0.
 \label{eq:S_free_energy_ratio}
\end{equation}
The spectral separation
$\alpha_{\rm s}-\alpha_{\rm f}>0$ determines the asymptotic rate of the
inversion, although the actual crossing time also depends on the mode
amplitudes and faster transient contributions.

Because the initial ordering satisfies
$\Delta F_\beta(\rho_{\rm h})
>\Delta F_\beta(\rho_{\rm c})$, while
Eq.~\eqref{eq:S_free_energy_ratio} gives the opposite ordering at
sufficiently long times, continuity guarantees at least one finite
crossing. More strongly, there exists a finite time
$t_{\rm per}$ such that
\begin{equation}
 \Delta F_\beta(\rho_{\rm h}(t))
 <
 \Delta F_\beta(\rho_{\rm c}(t)),
 \qquad
 t>t_{\rm per}.
 \label{eq:S_permanent_free_energy_inversion}
\end{equation}
Thus, eliminating the complete slow sector of the initially
higher-free-energy state, while retaining a nonzero slow component in
the comparison state, is sufficient to produce an eventual and
permanent thermodynamic Mpemba inversion.

When oscillatory modes produce several transient crossings,
$t_{\rm per}$ denotes the onset of the final permanently inverted
ordering and need not coincide with the first crossing time. In the
qutrit realization studied below, the curves cross only once, so
$t_{\rm per}$ coincides with the observed Mpemba time $t_{\rm M}$.

\begin{figure}[t]
\centering
\includegraphics[width=\columnwidth]{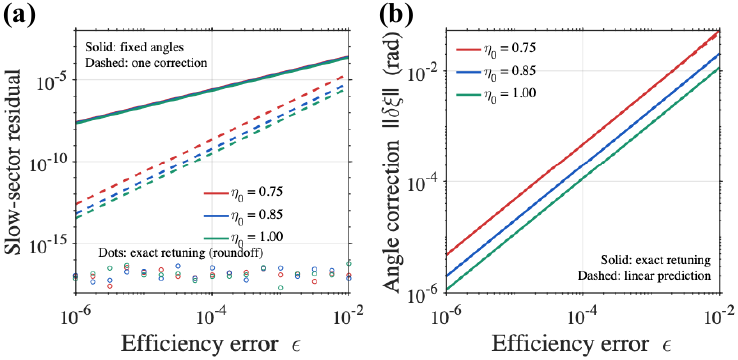}
\caption{
Restoring slow-sector cancellation after a detection-efficiency error.
Controls are initially tuned at $\eta_0=0.75$, $0.85$, and $1$,
while the actual common efficiency is $\eta=\eta_0-\epsilon$.
Left: Complete slow-sector residual
$\|\mathcal P_{\rm s}[\rho_{\rm h}-\tau_\beta]\|_{\rm HS}$
for fixed angles (solid), one first-order correction evaluated at
the nominal operating point (dashed), and exact retuning (circles).
The first two residuals scale as $\epsilon$ and $\epsilon^2$,
respectively; the exact-retuning residuals reflect numerical roundoff.
Right: Required angular correction
$\|(\delta\theta,\delta\phi)\|$, comparing exact retuning (solid)
with the linear prediction (dashed).
Phase differences are taken modulo $2\pi$.
Operating closer to $\eta_c$ requires larger corrections and reduces
the accuracy of the linear prediction at finite errors.
Other parameters are as in Fig.~\ref{fig:S_efficiency_relaxation}.
}
\label{fig:S_feedback_retuning}
\end{figure}

\section{Slow-mode response and feedback retuning}
\label{sec:S_slow_mode_response}

We now determine when a selective slow-sector cancellation can be
restored after small perturbations by retuning the feedback. Let
$\boldsymbol\xi\in\mathbb R^{n_{\rm a}}$ denote local coordinates on a
smooth branch of the Gibbs-preserving family $\mathcal G_\beta$.
Control variations leave the Hamiltonian, temperature, prescribed
initial states, and original jump operators fixed. We assume twice
continuously differentiable parameter dependence and work in a
neighborhood where the complete slow spectral sector remains isolated
and has a constant real dimension $d_{\rm s}$.

We begin with the spectral response to an admissible control
variation. Writing $\partial_\mu=\partial/\partial\xi_\mu$,
Eq.~\eqref{eq:S_feedback_generator} gives
\begin{equation}
 \partial_\mu\mathcal L_{\rm fb}[\rho]
 =
 \sum_k
 \left(
 \partial_\mu\mathcal F_k
 \right)
 [L_k\rho L_k^\dagger].
 \label{eq:S_generator_coordinate_derivative}
\end{equation}
For unitary feedback this response is an infinitesimal rotation,
$\partial_\mu\mathcal F_k[X]
=-i[G_{k\mu},\mathcal F_k[X]]$, with
$G_{k\mu}=i(\partial_\mu U_k)U_k^\dagger$ Hermitian.

For a diagonalizable controlled generator, choose biorthogonal
eigenoperators satisfying $(l_j|r_m)=\delta_{jm}$, where
$(A|B)=\operatorname{Tr}(A^\dagger B)$. Differentiating the eigenvalue
equation for a simple eigenvalue $\lambda_j$ and projecting onto
$l_j$ gives
\begin{equation}
 \partial_\mu\lambda_j
 =
 (l_j|\partial_\mu\mathcal L_{\rm fb}|r_j).
 \label{eq:S_eigenvalue_response}
\end{equation}
The real and imaginary parts describe changes in the decay rate and
oscillation frequency, respectively.

The same variation also changes the overlap of a fixed initial state
with the decay modes. At the evaluation point, choose the local
eigenoperator normalization such that
$(l_j|\partial_\mu r_j)=0$. Differentiating the left eigenvalue
equation then gives
$(\partial_\mu l_j|r_m)
=(l_j|\partial_\mu\mathcal L_{\rm fb}|r_m)/
(\lambda_j-\lambda_m)$ for $m\neq j$. Consequently, the amplitude
$a_{j,x}=(l_j|\rho_x-\tau_\beta)$ obeys
\begin{equation}
 \partial_\mu a_{j,x}
 =
 \sum_{m\neq j}
 \frac{
 (l_j|\partial_\mu\mathcal L_{\rm fb}|r_m)
 }{
 \lambda_j-\lambda_m
 }
 a_{m,x}.
 \label{eq:S_amplitude_response}
\end{equation}
The stationary contribution vanishes because
$a_{1,x}=\operatorname{Tr}(\rho_x-\tau_\beta)=0$.
Equations~\eqref{eq:S_eigenvalue_response} and
\eqref{eq:S_amplitude_response} separate spectral reordering from the
redistribution of a prescribed state's mode amplitudes. The
individual-mode formulas require a simple $\lambda_j$; the
projector-based criterion below also accommodates degeneracies within
an isolated slow sector.

We choose a smooth real frame for the Hermitian range of
$\mathcal P_{\rm s}$ and denote the coordinates of
$\mathcal P_{\rm s}[\rho_x-\tau_\beta]$ by
$\boldsymbol g_x\in\mathbb R^{d_{\rm s}}$. Consider an interior
selective cancellation point $\boldsymbol\xi_*$ satisfying
$\boldsymbol g_{\rm h}(\boldsymbol\xi_*)=\boldsymbol0$ and
$\boldsymbol g_{\rm c}(\boldsymbol\xi_*)\neq\boldsymbol0$. Define the
control matrix by
$[\mathsf C^{({\rm h})}]_{\alpha\mu}
=\left.\partial_\mu g_{{\rm h},\alpha}\right|_{\boldsymbol\xi_*}$.
An arbitrary infinitesimal slow-sector error $\boldsymbol e$ can be
cancelled precisely when
$\mathsf C^{({\rm h})}\delta\boldsymbol\xi=-\boldsymbol e$ is solvable
for every $\boldsymbol e\in\mathbb R^{d_{\rm s}}$. This is equivalent
to
\begin{equation}
 \operatorname{rank}\mathsf C^{({\rm h})}
 =
 d_{\rm s}.
 \label{eq:S_slow_control_rank}
\end{equation}
Thus, full row rank is necessary and sufficient for regular
first-order control of every slow-sector component. If the rank is
deficient, a nonzero left null vector identifies an error component
inaccessible to first-order retuning. This does not exclude a singular
cancellation or compensation of particular perturbations. At a
cancellation point, smooth changes of the slow-sector frame or
admissible control coordinates multiply the control matrix by
invertible matrices and therefore leave its rank unchanged.

To establish robustness under physical perturbations, let
$\boldsymbol p$ denote additional parameters and assume a smooth
admissible family satisfying
$\mathcal L(\boldsymbol\xi,\boldsymbol p)[\tau_\beta]=0$ throughout
the neighborhood considered. The initial states and $\tau_\beta$
remain fixed; any change in the Gibbs-preserving constraints must be
absorbed into this parametrization. At
$(\boldsymbol\xi_*,\boldsymbol p_0)$, define
$\mathsf B^{({\rm h})}
=\partial_{\boldsymbol p}\boldsymbol g_{\rm h}$. Taylor expansion gives
\begin{equation}
 \boldsymbol g_{\rm h}
 (\boldsymbol\xi_*+\delta\boldsymbol\xi,
  \boldsymbol p_0+\delta\boldsymbol p)
 =
 \mathsf C^{({\rm h})}\delta\boldsymbol\xi
 +
 \mathsf B^{({\rm h})}\delta\boldsymbol p
 +
 \mathcal O\!\left[
 \left(
 \|\delta\boldsymbol\xi\|+\|\delta\boldsymbol p\|
 \right)^2
 \right].
 \label{eq:S_slow_component_perturbation}
\end{equation}
Under Eq.~\eqref{eq:S_slow_control_rank}, the minimum-norm first-order
correction is
\begin{equation}
 \delta\boldsymbol\xi_{\min}
 =
 -
 \left(
 \mathsf C^{({\rm h})}
 \right)^+
 \mathsf B^{({\rm h})}\delta\boldsymbol p,
 \label{eq:S_minimum_control_retuning}
\end{equation}
where $+$ denotes the Moore--Penrose inverse. It removes the linear
residual, leaving a contribution of order $\|\delta\boldsymbol p\|^2$.
Other first-order corrections differ by an element of
$\ker\mathsf C^{({\rm h})}$. For fixed coordinate and normalization
conventions, the bound
$\|\delta\boldsymbol\xi_{\min}\|
\leq
\|\mathsf B^{({\rm h})}\delta\boldsymbol p\|/
\sigma_{\min}(\mathsf C^{({\rm h})})$
shows how a small singular value amplifies the required retuning.

Full row rank also guarantees exact local continuation. Choose
$d_{\rm s}$ controls corresponding to a nonsingular Jacobian minor and
hold the remaining controls fixed. The implicit-function theorem then
supplies nearby controls for which $\boldsymbol g_{\rm h}=\boldsymbol0$
after every sufficiently small admissible perturbation. For fixed
$\boldsymbol p$, the local solution set has dimension
$n_{\rm a}-d_{\rm s}$; it is locally isolated when
$n_{\rm a}=d_{\rm s}$. Continuity preserves the nonzero
comparison-state amplitude in a sufficiently small neighborhood. A
straight displacement in $\ker\mathsf C^{({\rm h})}$ preserves
cancellation only to first order; exact cancellation requires
following the solution set. These conclusions apply while the selected
cluster remains the complete slow sector. If the spectral ordering
changes, the slow projector and its control matrix must be reassessed.

Figure~\ref{fig:S_feedback_retuning} tests this construction in the
qutrit, where two feedback angles control the two real components of
the slow coherence pair. Controls are tuned at the nominal common
efficiency $\eta_0=0.75$, $0.85$, or $1$, and the actual efficiency is
$\eta=\eta_0-\epsilon$, with $10^{-6}\leq\epsilon\leq10^{-2}$. Common
detection loss preserves the Gibbs constraint, so the test isolates
the recovery of slow-sector cancellation. Fixed angles generate a
residual proportional to $\epsilon$, whereas one correction evaluated
at the nominal operating point reduces it to order $\epsilon^2$
[Fig.~\ref{fig:S_feedback_retuning}(a)]. Exact retuning restores
cancellation within numerical precision. The required angular
correction grows closer to the threshold, and the linear prediction
departs more visibly from exact retuning at finite errors
[Fig.~\ref{fig:S_feedback_retuning}(b)]. All perturbed efficiencies
shown remain above $\eta_c$.

\begin{figure}[t]
\centering
\includegraphics[width=\columnwidth]{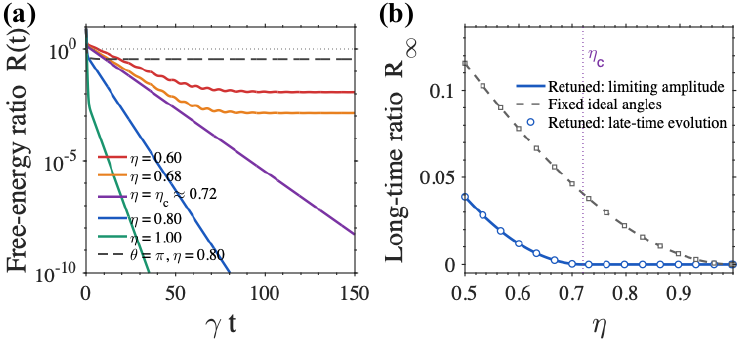}
\caption{
Finite-efficiency relaxation and permanent free-energy inversion.
Left: Free-energy ratio
$\mathcal R(t)=
\Delta F_\beta(\rho_{\rm h}(t))/
\Delta F_\beta(\rho_{\rm c}(t))$
for the indicated detection efficiencies, with feedback angles
retuned within the coherence-dominated regime.
Exact slow-sector cancellation gives $\mathcal R(t)\to0$ for
$\eta\geq\eta_c\simeq0.72$.
Below this threshold, a nonzero limiting ratio smaller than unity
still yields permanent inversion.
The horizontal dotted line marks $\mathcal R=1$; the gray dashed
curve shows the $\theta=\pi$, $\eta=0.80$ benchmark, whose limiting
ratio is $0.36$.
Right: Long-time ratio for retuned feedback (blue solid) and fixed
ideal-efficiency angles (gray dashed).
Lines give the slow-mode predictions, while circles and squares
show the corresponding late-time propagation results.
Parameters are $\beta\gamma=1$, $\omega_1/\gamma=1$,
$\omega_2/\gamma=1.5$, $\Delta p=0.03$,
$\chi_{\rm h}=0.012$, and $\chi_{\rm c}=0.020$.
}
\label{fig:S_efficiency_relaxation}
\end{figure}

\section{Finite-efficiency threshold and critical scaling}
\label{sec:S_efficiency_threshold}

For a common detection efficiency $\eta$, the controlled generator is
\begin{equation}
 \mathcal L_{\eta,\boldsymbol\xi}
 =
 \mathcal L_0
 +
 \eta\mathcal V_{\boldsymbol\xi},
 \qquad
 \mathcal V_{\boldsymbol\xi}
 =
 \sum_k
 \left[
 \mathcal F_k(\boldsymbol\xi)-\mathcal I
 \right]
 \mathcal J_k ,
 \label{eq:S_efficiency_generator}
\end{equation}
where $\boldsymbol\xi$ denotes admissible coordinates on the
Gibbs-preserving control set.  Since both
$\mathcal L_0[\tau_\beta]$ and
$\mathcal V_{\boldsymbol\xi}[\tau_\beta]$ vanish, common detection
loss weakens the feedback correction without changing the thermal
fixed point.  The qutrit exhibits two distinct thresholds:
$\eta_{\rm sp}$ marks the spectral selection of the coherence sector,
whereas $\eta_c$ marks the onset of its exact controllability.

\emph{Low-efficiency obstruction.---}
Let $\mathcal K_\beta$ be a compact admissible control domain and
assume that
$\boldsymbol\xi\mapsto\mathcal V_{\boldsymbol\xi}$ is continuous.
Suppose that the complete slow sector of $\mathcal L_0$ is isolated
and that the higher-free-energy state has a nonzero uncontrolled slow
overlap
$r_0=\|\mathcal P_{\rm s}^{(0)}
[\rho_{\rm h}-\tau_\beta]\|_{\rm HS}>0$.
Choose a closed contour $\Gamma$ enclosing this sector and no other
eigenvalues.  For sufficiently small $\eta$, the resolvent identity
gives
\begin{equation}
 \begin{aligned}
 \mathcal P_{\rm s}(\eta,\boldsymbol\xi)
 -
 \mathcal P_{\rm s}^{(0)}
 &=
 \frac{\eta}{2\pi i}
 \oint_\Gamma
 \mathcal R_{\eta,\boldsymbol\xi}(z)
 \mathcal V_{\boldsymbol\xi}
 \mathcal R_0(z)\,dz,
 \\
 \left\|
 \mathcal P_{\rm s}(\eta,\boldsymbol\xi)
 -
 \mathcal P_{\rm s}^{(0)}
 \right\|
 &\leq
 c_\Gamma\eta ,
 \end{aligned}
 \label{eq:S_projector_efficiency_bound}
\end{equation}
where
$\mathcal R_{\eta,\boldsymbol\xi}(z)
=(z-\mathcal L_{\eta,\boldsymbol\xi})^{-1}$ and
$c_\Gamma<\infty$ is uniform over
$\boldsymbol\xi\in\mathcal K_\beta$.

Writing
$\delta\rho_{\rm h}=\rho_{\rm h}-\tau_\beta$, the optimized residual
therefore satisfies
\begin{equation}
 \begin{aligned}
 R_{\rm h}(\eta)
 \equiv
 \min_{\boldsymbol\xi\in\mathcal K_\beta}
 \left\|
 \mathcal P_{\rm s}(\eta,\boldsymbol\xi)
 [\delta\rho_{\rm h}]
 \right\|_{\rm HS}
 \geq
 r_0
 -
 c_\Gamma\eta
 \|\delta\rho_{\rm h}\|_{\rm HS}.
 \end{aligned}
 \label{eq:S_low_efficiency_residual_bound}
\end{equation}
Hence $R_{\rm h}(\eta)$ remains strictly positive throughout a finite
interval above $\eta=0$.  Exact cancellation is therefore impossible
at sufficiently low efficiency whenever $r_0>0$.  This result proves
the existence of a positive low-efficiency obstruction but does not
determine the numerical value of $\eta_c$.

\emph{Continuation and fold scaling.---}
We now restrict attention to a region in which the complete slow
sector has fixed dimension and remains spectrally isolated.  Let
$\boldsymbol g_{\rm h}(\eta,\boldsymbol\xi)
\in\mathbb R^{d_{\rm s}}$
collect its independent real amplitudes.  At an exact cancellation
point, define the control matrix
$\mathsf C^{({\rm h})}
=\partial_{\boldsymbol\xi}\boldsymbol g_{\rm h}$
and the efficiency response
$\boldsymbol b_\eta=\partial_\eta\boldsymbol g_{\rm h}$.
If
$\operatorname{rank}\mathsf C^{({\rm h})}=d_{\rm s}$, the
implicit-function theorem guarantees a locally continued
exact-cancellation branch.  Its minimum-norm tangent obeys
\begin{equation}
 \mathsf C^{({\rm h})}
 \frac{d\boldsymbol\xi_*}{d\eta}
 =
 -
 \boldsymbol b_\eta,
 \qquad
 \left(
 \frac{d\boldsymbol\xi_*}{d\eta}
 \right)_{\!\min}
 =
 -
 \left(
 \mathsf C^{({\rm h})}
 \right)^+
 \boldsymbol b_\eta .
 \label{eq:S_efficiency_continuation}
\end{equation}
A regular interior branch can therefore terminate only through a loss
of slow-sector isolation, contact with a control boundary, or a loss
of full row rank.  The comparison-state amplitude must additionally
remain nonzero for the cancellation to represent a selective strong
Mpemba point.

For the square case
$n_{\rm a}=d_{\rm s}$ realized by the qutrit, suppose that the branch
terminates at
$(\eta_c,\boldsymbol\xi_c)$ with
$\operatorname{rank}\mathsf C_c^{({\rm h})}=d_{\rm s}-1$.
Let $\boldsymbol u$ and $\boldsymbol v$ be the corresponding left and
right null vectors.  If the projected efficiency response
$a=\boldsymbol u^{\mathsf T}
\partial_\eta\boldsymbol g_{\rm h}|_c$
and curvature
$b=\boldsymbol u^{\mathsf T}
D_{\boldsymbol\xi}^2\boldsymbol g_{\rm h}
[\boldsymbol v,\boldsymbol v]|_c$
are both nonzero, elimination of the nonsingular control directions
reduces the cancellation condition to the generic fold normal form
\begin{equation}
 0
 =
 a(\eta-\eta_c)
 +
 \frac{b}{2}q^2
 +
 \mathcal O\!\left[
 (\eta-\eta_c)^2
 +
 |\eta-\eta_c||q|
 +
 |q|^3
 \right],
 \label{eq:S_fold_normal_form}
\end{equation}
where $q$ is the control displacement along $\boldsymbol v$.

For the orientation realized in the qutrit, exact solutions lie on the
side $\eta>\eta_c$.  Equation~\eqref{eq:S_fold_normal_form} then gives
\begin{equation}
 \left\|
 \boldsymbol\xi_*(\eta)-\boldsymbol\xi_c
 \right\|,
 \;
 \sigma_{\min}
 \left(
 \mathsf C^{({\rm h})}
 \right)
 =
 \Theta\!\left(
 \sqrt{\eta-\eta_c}
 \right),
 \qquad
 R_{\rm h}(\eta)
 =
 \Theta(\eta_c-\eta)
 \quad
 (\eta<\eta_c).
 \label{eq:S_efficiency_critical_scaling}
\end{equation}
The vanishing singular value explains the increasingly ill-conditioned
retuning near $\eta_c$, while the linear residual describes the loss
of exact cancellation below threshold.  These scalings agree with
Fig.~3 of the main text.  The fold analysis applies near $\eta_c$,
where the slow sector is fixed, and is distinct from the lower
spectral-selection threshold $\eta_{\rm sp}$, at which the identity of
the complete slow sector changes.

Figure~\ref{fig:S_efficiency_relaxation} connects the cancellation
threshold to the actual free-energy relaxation.
For the protocols shown, the coherence pair remains the complete
slow sector, while the comparison-state amplitude is fixed at
$\chi_{\rm c}$.
Since both surviving coherence contributions enter the same
quadratic expansion of the relative entropy around $\tau_\beta$,
the long-time free-energy ratio is
$\mathcal R_\infty
=|a_{{\rm coh},{\rm h}}|^2/|\chi_{\rm c}|^2$.
This relation separates the effect of a reduced slow-mode weight
from that of its exact elimination.

At and above $\eta_c$, the retuned protocols eliminate the slow
amplitude of $\rho_{\rm h}$, and the free-energy ratio decays
exponentially to zero
[Fig.~\ref{fig:S_efficiency_relaxation}(a)].
Below threshold, the residual slow contribution instead produces
a nonzero plateau, approximately $1.2\times10^{-2}$ at $\eta=0.60$
and $1.4\times10^{-3}$ at $\eta=0.68$.
Both limits remain below unity, so permanent inversion survives
even though the two trajectories share the same asymptotic decay
exponent.
The $\theta=\pi$, $\eta=0.80$ benchmark provides another example:
its population-to-coherence source vanishes, but spectral
reordering leaves the limiting ratio
$(\chi_{\rm h}/\chi_{\rm c})^2=0.36$.
Thus, $\eta_c$ is the threshold for exact slow-sector elimination,
rather than a necessary threshold for permanent inversion.

Figure~\ref{fig:S_efficiency_relaxation}(b) compares the limiting
ratios predicted from the slow amplitudes with late-time
propagation.
Their agreement confirms the connection between the residual
projection and the asymptotic thermodynamic ordering.
Retuning restores a vanishing ratio for $\eta\geq\eta_c$,
whereas fixed ideal-efficiency angles leave a finite ratio for
$\eta<1$ throughout the range shown.
Near the fold, the linear scaling of the locally minimized
slow-sector residual further implies
$\mathcal R_\infty=\Theta[(\eta_c-\eta)^2]$
on the side without exact cancellation.
The free-energy ratio therefore approaches zero quadratically,
although the residual mode amplitude vanishes linearly.

\begin{figure}[t]
\centering
\includegraphics[width=\columnwidth]{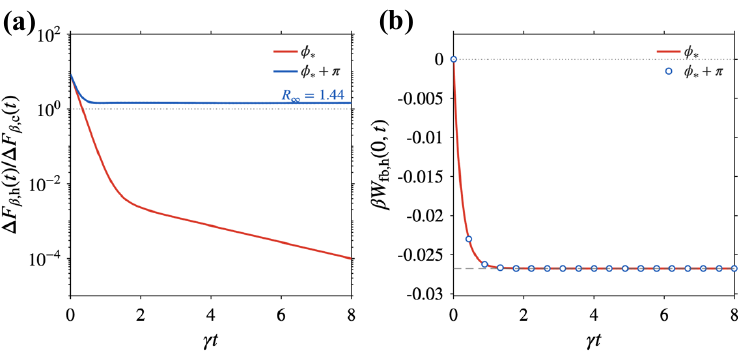}
\caption{Phase-dependent inversion at identical direct feedback work.
(a) Free-energy ratio
$R(t)=\Delta F_{\beta,{\rm h}}(t)/\Delta F_{\beta,{\rm c}}(t)$
for the optimized phase $\phi_*$ (red) and $\phi_*+\pi$ (blue).
Their asymptotic ratios are zero and $1.44$, respectively.
The dotted line marks $R=1$.
(b) Cumulative direct work for $\rho_{\rm h}$.
The red curve and blue circles coincide; the dashed line marks
$\beta W_{{\rm fb},{\rm h}}(0,\infty)\simeq-0.0268$.
Both protocols use $\eta=1$ and $\theta=\theta_*\simeq0.7872\pi$,
with $\phi_*\simeq1.9037\pi$.
Other parameters are $\beta\gamma=1$, $\omega_1/\gamma=1$,
$\omega_2/\gamma=1.5$, $\Delta p=0.03$,
$\chi_{\rm h}=0.012$, and $\chi_{\rm c}=0.020$.
}
\label{fig:S_phase_work}
\end{figure}

\section{Direct feedback work}
\label{sec:S_feedback_work}

We now quantify the energy exchanged directly between the feedback
pulses and the system. The pulses implement instantaneous unitary
operations $\mathcal F_k[X]=U_kXU_k^\dagger$ and begin and end with
the same system Hamiltonian $H$. When energy exchange with the
reservoir during each pulse is negligible, the pulse-induced change
of the system energy is identified as work. For a general nonunitary
feedback map this change need not be entirely work. Throughout each
relaxation run, the feedback parameters and detection efficiencies
are held fixed.

A jump in channel $k$ occurs at the rate
$r_k(t)=\operatorname{Tr}[L_k\rho(t)L_k^\dagger]$ and prepares
$\sigma_k(t)=L_k\rho(t)L_k^\dagger/r_k(t)$ whenever $r_k(t)>0$. The
conditional pulse work is
$w_k(t)=\operatorname{Tr}[(U_k^\dagger HU_k-H)\sigma_k(t)]$, with
positive work denoting energy supplied to the system. Averaging over
detected jumps gives
\begin{equation}
 \dot W_{\rm fb}(t)
 =
 \sum_k\eta_k
 \operatorname{Tr}
 \left[
 \left(U_k^\dagger HU_k-H\right)
 L_k\rho(t)L_k^\dagger
 \right].
 \label{eq:S_feedback_work_rate}
\end{equation}
The jump rate cancels the normalization of $\sigma_k(t)$, so the
conditional work enters the average without additional weighting. A
channel with $[U_k,H]=0$ contributes no direct work.

The reservoir-induced energy exchange is already contained in
$\mathcal L_0$. Defining
$E(t)=\operatorname{Tr}[H\rho(t)]$ and
$\dot Q_0(t)=\operatorname{Tr}\{H\mathcal L_0[\rho(t)]\}$, the
feedback master equation gives
\begin{equation}
 \frac{dE(t)}{dt}
 =
 \dot Q_0(t)+\dot W_{\rm fb}(t).
 \label{eq:S_system_first_law_feedback}
\end{equation}
The heat current is evaluated on the evolving controlled state.
Free-energy contraction constrains the combined changes of energy and
entropy, but it does not fix the sign of the instantaneous feedback
power.

Taking the energy moment of the efficiency-weighted Gibbs-preserving
condition yields
\begin{equation}
 \left.\dot W_{\rm fb}\right|_{\rho=\tau_\beta}=0.
 \label{eq:S_equilibrium_feedback_work}
\end{equation}
This is again a collective cancellation: individual channels may
exchange nonzero energies at equilibrium, yet their average powers
sum to zero.

For either initial preparation, the accumulated signed pulse work is
\begin{equation}
 W_{\rm fb}^{(\nu)}(0,t)
 =
 \int_0^t
 \dot W_{\rm fb}^{(\nu)}(s)\,ds,
 \label{eq:S_accumulated_feedback_work}
\end{equation}
with $\nu\in\{{\rm h},{\rm c}\}$. Evaluating this expression at
$t=t_{\rm M}$ gives the direct work exchanged before the free-energy
crossing.

For the qutrit, the integral admits a simple exact expression. The
two emission jumps prepare $P_0$ and $P_1$, respectively, so the
corresponding conditional pulse works are
$w_1=\omega_1\sin^2(\theta/2)$ and
$w_2=-\omega_1\sin^2(\theta/2)$. With a common detection efficiency
and $x=\eta\sin^2(\theta/2)$, the total feedback power becomes
\begin{equation}
 \dot W_{\rm fb}(t)
 =
 \omega_1 x
 \left[
 \gamma_1^-p_1(t)-\gamma_2^-p_2(t)
 \right]
 =
 \omega_1 x\,\dot p_2(t).
 \label{eq:S_qutrit_feedback_power}
\end{equation}
The last equality uses
$\dot p_2=\gamma_2^+p_1-\gamma_2^-p_2$ together with the chosen rate
relation $\gamma_2^+=\gamma_1^-=\gamma$. Since $x$ is constant during
each run,
\begin{equation}
 W_{\rm fb}^{(\nu)}(0,t)
 =
 \omega_1 x
 \left[
 p_{2,\nu}(t)-p_{2,\nu}(0)
 \right].
 \label{eq:S_qutrit_accumulated_work}
\end{equation}
The accumulated direct work therefore follows from the population
trajectory alone, without a separate power integration. Although the
phase $\phi$ controls interference in the coherence amplitude, it
leaves the population dynamics and the direct work unchanged at fixed
$(\eta,\theta)$.

The comparison state has Gibbs populations, which remain unchanged
because the population block is closed. Hence
$W_{\rm fb}^{({\rm c})}(0,t)=0$ for all $t$. For the
higher-free-energy state,
$p_{2,{\rm h}}(0)=p_2^\beta+\Delta p$, and relaxation to the Gibbs
state gives
$W_{\rm fb}^{({\rm h})}(0,\infty)=-\omega_1x\Delta p$. For
$\Delta p>0$ and $x>0$, the pulses extract net energy over the
complete relaxation. The
work accumulated before the crossing instead follows from
Eq.~\eqref{eq:S_qutrit_accumulated_work} evaluated at $t_{\rm M}$.

These signed quantities describe direct pulse--system energy
exchange. They exclude the resources required for detection, pulse
generation, record processing, and controller reset. Neither a
negative accumulated pulse work nor a vanishing equilibrium power
determines the total operating cost of the feedback apparatus.

Figure~\ref{fig:S_phase_work} compares two Gibbs-preserving protocols
that differ only in the feedback phase, with the same mixing angle,
detection efficiency, and initial states. The phase leaves the
population dynamics and Liouvillian eigenvalues unchanged, but it
changes the interference between the initial and population-induced
contributions to the slow coherence amplitude. At $\phi_*$, the
latter contribution is $-\chi_{\rm h}$ and cancels the initial
amplitude. Shifting the phase by $\pi$ reverses its sign, giving a
total slow amplitude $2\chi_{\rm h}$. The comparison state retains
the amplitude $\chi_{\rm c}$ in both protocols. Consequently, the
optimized ratio tends to zero, whereas the phase-shifted ratio
approaches $R_\infty=(2\chi_{\rm h}/\chi_{\rm c})^2=1.44>1$
[Fig.~\ref{fig:S_phase_work}(a)]. The phase shift thus removes the
permanent inversion even though the decay spectrum is unchanged.

The two protocols nevertheless exchange exactly the same cumulative
direct work at every common evolution time
[Fig.~\ref{fig:S_phase_work}(b)]. Their shared population trajectory
gives
$W_{{\rm fb},{\rm h}}(0,t)
=\omega_1x[p_{2,{\rm h}}(t)-p_{2,{\rm h}}(0)]$,
where $x=\sin^2(\theta_*/2)$ at ideal detection, and hence
$\beta W_{{\rm fb},{\rm h}}(0,\infty)
=-\beta\omega_1x\Delta p\simeq-0.0268$. Both protocols extract the
same net pulse work from the system. Their system energies also
coincide at equal times, so the different free energies reflect
different entropy trajectories. This comparison isolates the role of
slow-mode interference in the inversion: identical direct energy
exchange does not determine the asymptotic free-energy ordering.

\section{Requirements for an experimental implementation}
\label{sec:S_experiment}

A superconducting circuit provides a possible setting for the qutrit
protocol. Coherent control and state tomography of the three lowest
transmon levels have been demonstrated~\cite{Bianchetti2010}. In our
notation, the adjacent transitions have frequencies $\omega_1$ and
$\omega_2$, while $\omega_1+\omega_2$ is the energy of $|2\rangle$
relative to $|0\rangle$, with $\hbar=1$. A selective microwave pulse
on the $|0\rangle$--$|1\rangle$ transition implements
$U_{\theta,\phi}$, with its envelope and phase determining the
rotation angle and axis. Pulse shaping must suppress leakage outside
the intended subspace.

Realizing the dissipative dynamics requires control of both absorption
and emission. The rates must satisfy
$\gamma_i^+/\gamma_i^-=e^{-\beta\omega_i}$ together with the
equilibrium activity matching
$\gamma_1^-p_1^\beta=\gamma_2^-p_2^\beta$.
Quantum bath engineering in superconducting circuits provides a
possible starting point. The illustrative parameters
$\omega_2/\omega_1=1.5$ and $\omega_1/\gamma=1$ used here, however,
do not describe the bare spectrum and relaxation of a conventional
transmon. A device implementation must
either reestablish the cancellation for its measured parameters or
specify an engineered effective Hamiltonian and dissipative channels.
In the latter case, the Gibbs reference must be verified for that
effective Hamiltonian.

The monitoring must reproduce the emission-counting dynamics assumed
in the model. The efficiency $\eta$ is the probability that an
emission in either monitored channel produces a valid feedback
trigger, including photon collection, transmission, detection, and
detector availability. It is distinct from the fidelity of a
single-shot measurement of the transmon state. Continuous dispersive
state readout would also introduce measurement-induced dephasing,
which changes the coherence dynamics. Microwave photon counters
demonstrate the relevant detection capability~\cite{Lescanne2020}; their operating bandwidth, dead time, and total
collection efficiency must be evaluated for both emission channels.
The threshold $\eta_c\simeq0.72$ applies to the parameters studied
here under common, independent detection losses. It must be
recomputed if the Hamiltonian, rates, or prescribed initial states
are changed. False triggers require a separate rate calibration and
are not represented by $1-\eta$.

The feedback latency includes photon detection, signal processing,
and pulse application. A loop delay of $250$\,ns was reported in a
superconducting-qubit feedback experiment~\cite{Vijay2012},
demonstrating fast electronic control; the full latency of a photon
counting implementation must be measured separately. The short-delay
regime requires $\Gamma_{\rm d}\tau_{\rm loop}\ll1$, where
$\Gamma_{\rm d}$ bounds the relevant dissipative rates, with coherent
phase evolution accounted for in the pulse timing. Finite delay
requires dynamics that retain the pending feedback record and can
modify both the stationary state and the slow-mode amplitudes.
Gibbs preservation and selective cancellation must therefore be
checked with the measured response, including additional dephasing
and leakage. A small delay alone does not establish exact
slow-sector elimination.

An experimental test would first verify that the Gibbs state remains
stationary with feedback enabled. The same two initial states would
then be prepared repeatedly, and tomography at selected evolution
times would reconstruct the unconditional states and their
free-energy excesses. The crossing and the subsequent decay rates
could be compared with the no-feedback trajectories and repeated
after retuning at different calibrated efficiencies. This directly
tests the collective balance and selective relaxation on which the
protocol relies.

\end{document}